
\font\subtit=cmr12
\font\name=cmr8
\def\ng{\backslash\!\!\!\!D}

\def\hyp{{{\rm\bf H}^2}}
\def\eucl{{{\rm\bf R}^{n-2}}}

\def\sph{{{\rm\bf S}^2}}
\def\dsk{{{\rm\bf D}^2}}
\def\fot#1{F(1-\beta,\beta;1|#1)}

\def\mti{g^{z\bar z}}
\def\xiv{\vec\xi}
\input harvmac

\def\plb#1#2#3#4{#1, {\it Phys. Lett.} {\bf {#2}}B (#3), #4}
\def\npb#1#2#3#4{#1, {\it Nucl. Phys.} {\bf B{#2}} (#3), #4}
\def\prl#1#2#3#4{#1, {\it Phys. Rev. Lett.} {\bf {#2}} (#3), #4}
\def\prd#1#2#3#4{#1, {\it Phys. Rev.} D{\bf {#2}} (#3), #4}
\def\cmp#1#2#3#4{#1, {\it Comm. Math. Phys.} {\bf {#2}} (#3), #4}
\def\lmp#1#2#3#4{#1, {\it Lett. Math. Phys.} {\bf {#2}} (#3), #4}
\def\ap#1#2#3#4{#1, {\it Ann. Phys.} {\bf {#2}} (#3), #4}

\def\ijmpa#1#2#3#4{#1, {\it Int. Jour. Mod. Phys.} {\bf A{#2}} (#3), #4}

\def\cqg#1#2#3#4{#1, {\it Class. Q. Grav.} {\bf {#2}} (#3), #4}
\def\Fort#1#2#3#4{#1, {\it Fortsch. Phys.} {\bf {#2}} (#3), #4}
\def\UTN#1#2#3
{\TITLE{UTF-#1-\number\yearltd}
{#2}{#3}}
\def\TITLE#1#2#3{\nopagenumbers\abstractfont\hsize=\hstitle\rightline{#1}
\vskip 1in
\centerline{\subtit #2}
\vskip 1pt
\centerline{\subtit #3}\abstractfont\vskip .5in\pageno=0}%
\UTN{344}
{FIELD THEORIES ON THE POINCAR\'E DISK}{}

\centerline{F{\name RANCO} F{\name ERRARI}}\smallskip
\centerline{\it Dipartimento di Fisica, Universit\'a di Trento,
38050 Povo (TN), Italy}
\centerline{\it and INFN, Gruppo Collegato di Trento, Italy}
\vskip 4cm
\centerline{ABSTRACT}
\vskip 1cm
{\narrower
The massive scalar field theory and the
chiral Schwinger model are quantized
on a Poincar\'e disk of radius $\rho$.
The amplitudes are derived
in terms of hypergeometric functions. The behavior at long distances and
near the boundary of some of the
relevant correlation functions is studied. The exact computation of the chiral
determinant appearing in the Schwinger model is obtained exploiting
perturbation theory. This calculation poses interesting mathematical
problems, as the Poincar\'e disk is a noncompact manifold with a metric
tensor which diverges approaching the boundary. The results presented in this
paper are very useful in view of possible extensions to general
Riemann surfaces. Moreover, they could also shed some light in the quantization
of field theories on manifolds with constant curvature scalars in higher
dimensions.
}
\Date{February 1995}
\newsec {INTRODUCTION}
\vskip 1cm
This paper treats the quantization of field theories on a two dimensional
manifold with constant curvature scalar $R$. Positive values of $R$
correspond to the topology of a sphere, whereas negative values
correspond to hyperbolic geometries, like the Poincar\'e upper half
plane or the Poincar\'e disk.
Particular attention is devoted to the
case of negative curvatures, which is relevant in many
different contexts \ref\calwil{\npb{C. G. Callan and F.
Wilczek}{340}{1990} {366}.}\nref\cvz{\prd{G. Cognola, K. Kirsten,
L. Vanzo and S.
Zerbini} {49}{1994}{5307};\plb{A. A. Bytsenko, S. D. Odintsov and S. Zerbini}
{336}{1994}{355}; \prd{G. Cognola and
L. Vanzo}{47}{1993}{4575}.}\nref\zerb{\lmp{S. Zerbini}{27}
{1993}{19}.}\nref\pbt{\cmp{J.
Palmer, M. Beatty and C. A. Tracy}
{165}{1994}{97}; \npb{R. Narayanan and C. A.
Tracy}{340}{1990}{568}.}\nref\cfthyp{\prl{P. Kleban and I.
Vassilieva}{72}{1994}{3929}; \ijmpa{Z. Haba}{4}{1989}
{267}; \ap{A. Comtet}{173}{1987}{173}.}--\ref\grosche{\Fort{C.
Grosche}{38}{1990}{531};
\cmp{J. Bolte and C. Grosche}{163}{1994}{217}.}.
\smallskip
In the first part of the paper we deal with massive scalar fields
coupled to $R$ through a coupling constant $\lambda$. This model is
interesting in itself. After a field redefinition, for example, the
free equations of motion on the Poincar\'e upper half plane are
equivalent to the Euler-Poisson-Darboux equations describing the
propagation of waves in a polytropic gas. Moreover, since the
manifolds with constant curvature are harmonic, it is always possible
to reduce the equations of motion of the fields (also in the presence
of self-interactions) to ordinary differential equations of the second
order \ref\fried{F. G. Friedlander, The wave Equation on a Curved Space-Time,
Cambridge University Press, Cambridge, London, New York, Melbourne
1975.}.
This is a valid alternative to the heat kernel techniques in computing
the correlation functions.
The free propagator is for instance solution of an equation of the
hypergeometric type \foot{We notice that similar equations have been
found in ref. \pbt\ for the massive fermionic fields.}, independently
of the fact that $R$ is positive or negative.
In this sense it is possible to unify within the approach presented
here the treatment of scalar fields on different topologies like the
Poincar\'e disk or the sphere. The analogies between the two cases are
however fortuitous and exist only at a formal level. As a matter of
fact, the sphere is a manifold without boundary, whereas on the
Poincar\'e disk or upper half plane proper boundary conditions should
be imposed\foot{Let us remember that the in both cases of the disk
and of the upper half plane the boundary does not belong to the
manifold. Nevertheless, it is necessary to specify the limiting
conditions
with which the fields approach the boundary.}.
The latter are however determined by the geometry. For instance we
show that, once we require that the propagator has only the physical
singularity in the origin, automatically Dirichlet boundary conditions
are chosen, in agreement with ref. \calwil.
\smallskip
In the second part of the paper we deal with the Schwinger model
\ref\schw{J. Schwinger, {\it Phys. Rev.} {\bf 128} (1962), 2425.} on a
Poincar\'e disk of radius $\rho$.
Due to its physical relevance
\ref\flat{S. Coleman, {\it Phys. Rev.} {\bf D11} (1975), 3026;
S. Donaldson, {\it J. Diff. Geom.}
{\bf 18} (1983), 269;
R. Jackiw and R. Rajaraman, {\it Phys. Rev. Lett.} {\bf 54} (1985), 1219;
L. Faddev and S. Shatashvili, {\it Phys. Lett.} {\bf 183B} (1987), 311;
T. P. Killingback, {\it Phys. Lett.} {\bf 223B} (1989) 357;
A. I. Bocharek and M. E. Schaposhnik, {\it Mod. Phys. Lett.} {\bf A2} (1987),
991; J. Kripfganz and A. Ringwald, {\it Mod. Phys. Lett.} {\bf A5}
(1990), 675.}\nref\cmpl{J. H. Lowenstein, J. A. Swieca,
{\it Ann. Phys.} {\bf 68}
(1961), 172;
A. Z. Capri and R. Ferrari, {\it Nuovo Cim.} {\bf 62A} (1981), 273;
{\it Journ. Math. Phys.} {\bf 25} (1983), 141;
G. Morchio, D. Pierotti and F. Strocchi, {\it Ann. Phys.} {\bf 188}
(1988), 217;
A. K. Raina and G. Wanders,
{\it Ann. of Phys.} {\bf 132} (1981),
404.}\nref\aar{E. Abdalla, M. C. B. Abdalla and K. Rothe, {\it Nonperturbative
Methods in 2 Dimensional Quantum Field Theory}, World Scientific, Singapore,
1991.}--\ref\manton{N. Manton,
{\it Ann. Phys.} {\bf 159} (1985), 220; J. E.
Hetrick and Y. Hosotani, {\it Phys. Rev.} {\bf 38} (1988), 2621.}, the
Schwinger model and its generalizations
have been quantized on many different topologies
\ref\jaw{C.
Jayewardena, {\it Helv. Phys. Acta} {\bf 61} (1988),
636.}\nref\sphere{A. Bassetto and L. Griguolo, {\it The Generalized Chiral
Schwinger Model on the Two Sphere}, Preprint DFPD-94-TH-62,
hep-th/9411229.}\nref\Joos{H. Joos
and S. I. Azakov, {\it The Geometric Schwinger Model on
the Torus. 2.}, Preprint DESY-94-142, H. Joos,
{\it Nucl. phys.} {\bf B17}
(Proc. Suppl.) (1990), 704; {\it Helv. Phys. Acta} {\bf 63} (1990),
670.}\nref\rajeev{S. G. Rajeev, {\it Phys. Lett.} {\bf 212B} (1988),
203, K. S. Gupta,
R. J. Henderson, S. G. Rajeev and O. T. Turgut, {\it Jour. Math. Phys.}
{\bf 35} (1994), 3845.}\nref\wipf{I. Sachs and A. Wipf,
{\it Helv. Phys. Acta} {\bf 65} (1992), 653;
{\it Phys. Lett.} B{\bf 326}
(1994), 105.}\nref\netodas{\prd{J. Barcelos-Neto and A. Das}{33} {1986}
{2262}.}\nref\ferqed{F. Ferrari, {\it Class. Q. Grav.}
{\bf 10} (1993), 1065; {\it Helv. Phys. Acta} {\bf 67} (1994),
702.}--\ref\ffscht{
F. Ferrari, {\it On the Schwinger Model on Riemann Surfaces}, to
appear in {\it Nucl. Phys.} {\bf B}, hep-th/9310055.}
but not, until now, on an hyperbolic two dimensional manifold.
\smallskip
Conceptually, the computation of the anomaly in the presence of negative
curvature presents many difficulties.
The reason is that the Poincar\'e disk, like the upper half plane, is
limited by a boundary. The latter does not belong to the manifold,
however the behavior of the fields as they approach the boundary must be given.
Moreover, the metric tensor becomes singular exactly on the boundary.
Most of the mathematical and physical literature discussing the calculation of
chiral determinants on curved space-times
\ref\chiramath{M. F. Atiyah, V. K. Patodi and I. M. Singer, {\it Math.
Proc. Camb. Phil. Soc.}{\bf 77} (1975), 43;
M. F. Atiyah, R. Bott and V. K. Patodi, {\it Inventiones Math.} {\bf
19} (1973), 279; R. T. Seeley, {\it Am.
Math. Soc. Proc. Symp. Pure Math.} {\bf 10} (1967), 288; P. B. Gilkey,
{\it Proc. Symp. Pure Math.} {\bf 27} (1973); {\it Jour. Diff. Geom.}
{\bf 10} (1975), 601.}\nref\chiraphys{\cmp{S. W. Hawking}{55}{1977}{133};
\cqg{D. M. McAvity and H. Osborn}{8}{1991}{603}; \plb{H.
Leutwyler}{152}{1985}{78}; \cmp{R. E. Gamboa Saravi, M. A. Muschietti and
J. Solomin}{89}{1983}{363}; \npb{L. Alvarez-Gaum\'e and P.
Ginsparg}{243} {1984}{449}; \cmp{O. Alvarez, I. M. Singer and B.
Zumino} {96}{1984}{409}; \prd{L. Bonora, M. Bregola and P. Pasti}
{31}{1985}{2645}; \npb{A. Andrianov and L. Bonora}{233}{1984}{232}; {\it
ibid.} 247; \plb{L. Bonora, P. Cotta-Ramusino and C. Reina}{126}{1983}
{305}.}--\ref\msit{\ap{A. V. Mishchenko and Yu. A.
Sitenko}{218}{1992}{199}}
does not treat this
particular situation explicitly.
An exception is however provided by ref. \ref\niesem{\npb{A. J. Niemi
and G. W. Semenoff}{269}{1986}{131}.}. Fortunately, in order to derive
the form of the anomaly in the present case one can also exploit a
perturbative approach, avoiding mathematical complications \aar.
This strategy will be adopted here, showing that there are no terms
related to the boundary in the chiral determinant. Finally, in order
to study the behavior of the correlation functions of the Schwinger
model at short and long distances, we will use the propagator derived
for the massive scalar fields in the first part of the paper. This
task is made relatively easy by the fact that, as already remarked,
this propagator is expressed in terms of hypergeometric functions. The
2--point $\bar\psi\psi$ correlator is thoroughly studied in this way.
\smallskip
The presentation of the above discussed results is organized as
follows. In Section 2 the theory of massive scalar fields coupled with
the constant curvature scalar is quantized on the Poincar\'e upper
half plane. The propagator is derived solving an hypergeometric
equation. The behavior of the propagator and the choice of the
boundary conditions are discussed in details. In Sections 3 and 4 are
treated the
cases of the sphere and of the Poicarer\'e disk respectively. The
Poincar\'e disk is equivalent to the upper half plane up to a
conformal transformation. Finally, the Schwinger model on a Poincar\'e
disk of radius $\rho$ is investigated in Section 5.
The fermionic propagator and the chiral determinants are computed using a
perturbative approach. After bosonization, the effective theory
becomes as in the flat case a free field theory of massive mesons and of free
massless fermions. At the end of the Section the behavior of the
correlation functions and in particular of the $\bar\psi\psi$ 2--point
function are studied in details. All the formulas concerning
hypergeometric functions used in this paper have been listed in
Appendix A. In Appendix A we also show that the propagators of the
scalar fields with the proper boundary conditions have only the
physical singularity at the origin.
\vskip 1cm
\newsec{MASSIVE SCALAR FIELDS ON $\hyp$}
\vskip 1cm
Let us consider the functional:
\eqn\action{
S=\int_\hyp d^2x\sqrt{g}\left({1\over 2}\partial_\mu\varphi\partial^\mu
\varphi+{\mu^2\over 2}\varphi^2+(\lambda+1)R\varphi\right)}
This action describes a massive scalar field theory on the upper half plane
$\hyp$  parametrized
by the coordinates $x,y$. The fields are
coupled with the
curvature scalar $R$ through the
real parameter $\lambda$.
The metric on $\hyp$ is Euclidean:
\eqn\metr{g_{\mu\nu}={\rm diag}\left({1\over y^2},{1\over y^2}\right)}
Starting from this metric it is easy to see that the curvature scalar
$R$ amounts to a negative constant, which can be renormalized in such
a way that $R=-1$.
\smallskip
At this point, it is convenient to introduce on $\hyp$
the complex
coordinates $z=x+iy$ and $\bar z=x-iy$.
Then the equation of motion ${\delta S\over\delta\varphi}=0$
for the propagator $G_\hyp(z,w)\equiv\langle\tilde\varphi(z,\bar z)
\tilde\varphi(w,\bar w)\rangle$
reads as follows:
\eqn\eqsmot{\left[(z-\bar z)^2\partial_z\partial_{\bar z}+\mu^2\right]
G_\hyp(z,w)=\delta^{(2)}(z,w)-(\lambda+1)R}
Here we have exploited the identity $\partial_z\partial_{\bar z}={1\over 4}
(\partial_1^2+\partial_2^2)$ and the fact that, in complex coordinates, the
components of the metric on $\hyp$ are $g^{z\bar z}=g^{\bar zz}=
({\rm Im}[z])^2$,
where $({\rm Im}[z])^2=-{1\over 4}(z-\bar z)^2$.
If $\lambda=-1$, eq. \eqsmot\ becomes the usual equation
$$\left[-\triangle+\mu^2\right] G_\hyp(z,w)=\delta^{(2)}(z,w)$$
It is interesting to notice that if $\lambda=1$
eq. \eqsmot\ is equivalent to the
Euler-Poisson-Darboux equation
describing the propagation of waves in a polytropic gas.
After the change of variables:
\eqn\changevar{
\left\{\matrix{u=z&\bar u=-\bar z\cr v=w&\bar v=-\bar w\cr}\right.
\qquad\qquad\qquad
G_\hyp(u,v)=\left({u+\bar u\over v+\bar v}\right)^\beta f(u,v)}
eq. \eqsmot\ becomes in fact
the Euler-Poisson-Darboux equation:
\eqn\eulpois{\partial_u\partial_{\bar u}f+{\beta\over u+\bar u}\left(
\partial_uf+\partial_{\bar u}f\right)=0}
where $\beta$ is defined by the relation: $\beta(\beta-1)=\mu^2$.
\smallskip
To solve \eqsmot\ we use the fact that a space with constant curvature is
harmonic.
Denoting with $\Gamma$
the square of the geodetic distance between
two points on $\hyp$, this means that
$\triangle\Gamma$ is a function of $\Gamma$.
As a consequence, since the propagator must be
a function of $\Gamma$, the equation of motion \eqsmot\ becomes
an ordinary differential equation of the second order in $\Gamma$ \fried.
\smallskip
In our case, it will be more convenient to choose instead of $\Gamma$
the anharmonic ratio
\eqn\anhrat{X\equiv-{(z-w)(\bar z-\bar w)\over
(z-\bar z)(w-\bar w)}}
so that $G_\hyp(z,w)\equiv G_\hyp(X)$, with $X\ge 0$ by construction.
This is possible because the geodetic length $\Gamma$ is a function of $X$.
Considering now a generic function $F(X)$ of $X$ as an implicit
function of $z$ and $\bar z$, it is straightforward to
prove the following relation:
\eqn\relfund{(z-\bar z)^2\partial_z\partial_{\bar z}F(X)=
-X(1+X)F^{\prime\prime}(X)-(1+2X)F'(X)- XF'(X)\delta^{(2)}(z,w)}
where the prime denotes the derivative in $X$ and the Dirac
delta function is
defined as usual by:
$$\delta^{(2)}(z,w)=-{1\over 4\pi}
g^{z\bar z}\partial_z\partial_{\bar z}{\rm log}(X)+{R\over 4\pi}$$
It is clear from the last term in eq. \relfund\ that, in order to generate
a Green function with the correct singularity
in $z=w$, the behavior of $F(X)$ should be logarithmic when
$X\sim 0$, so that $XF'(X)=1+O(X)$.
Substituting eq. \relfund\ in eq. \eqsmot, we get a
second order differential equation in $G_\hyp(X)$ of the kind:
\eqn\prehyp{-X(X+1)G^{\prime\prime}_\hyp(X)-
(1+2X)G'_\hyp(X)+\mu^2G_\hyp(X)=-(\lambda+1)R}
This is the desired final expression of the equations of motion.
\smallskip
At this point, we notice that
eq. \prehyp\ is hypergeometric\foot{The properties of the
hypergeometric functions that will be used here can be found for
example in \ref\kamke{E. Kamke, {\it Differentialgleichungen, L\"osungmethoden
und L\"osungen}, I {\it (Gew\"ohnliche Differentialgleichungen)} (7th ed.),
Akademische Verlagsgesellschaft, Leipzig (1961).}--\ref\erdelyi{A. Erd\'elyi
and H. Bateman, {\it Higher trascendental functions}, Vol. I, Mc Graw-Hill
Book Company Inc., New York 1953.}.}.
As a matter of fact, substituting $y=-X$ and $\eta(y)=G_\hyp(X)$, eq. \prehyp\
becomes:
\eqn\hyperg{y(1-y)\eta^{\prime\prime}+(1-2y)\eta'+\mu^2\eta=-(\lambda+1)R}
Now it is easy to construct
the solutions of eq. \hyperg.
The only problem is to choose the physical boundary
conditions when $z$ and $w$ approach the boundary of $\hyp$ on the real line.
Setting $\beta={1\over 2}-\sqrt{{1\over 4}+\mu^2}$, we have that
\eqn\yone{\eta_1(y)= \fot{y}}
is one of the two independent solutions of eq. \hyperg\ in the homogeneous
case, i.e. when $\lambda=1$.
Here $F(a,b;c|z)$ denotes the hypergeometric function
\eqn\serepr{F(a,b;c|z)=\sum\limits_{k=0}^\infty{(a)_k(b)_k\over (c)_k
k!}z^k}
with $(a)_k=a(a+1)(a+2)\ldots (a+k-1)$.
Clearly, $\eta_1(y)$ is not singular in $y=0$ but has a logarithmic
divergence in $y=1$.
The other independent solution of eq. \hyperg\ when $\lambda=-1$ is given by:
\eqn\ytwo{\eta_2(y)=\fot{y}\int^y_{C_1}{[t(1-t)]^{-1}\over [\fot{t}]^2}}
$\eta_2(y)$ has the desired logarithmic singularity in $y=0$ with the correct
sign, but diverges also in $y=1$ due to the presence of the
factor $\fot{y}$. To avoid this problem, it is sufficient to take
$C_1=1$ in eq. \ytwo.
This choice can still be potentially dangerous, because apparently the
integrand
has a simple pole at $t=1$. However, this singularity
is integrable because of the denominator, which has the behavior
$[\fot{t}]^2\sim{\rm log}^2(t)$ in a neighborhood of the point $t=1$.
Having the two independent solutions of the homogeneous problem,
the final ingredient
in order to solve eq.
\hyperg\ completely is a particular solution.
This is given by:
\eqn\ypart{\bar \eta(y)=-R(\lambda+1)\fot{y}\int^y_{C_1}
dt{[t(1-t)]^{-1}\over[\fot{t}]^2}
\int^t_{C_2}ds\fot{s}}
Again, taking $C_1=1$ in the above equation,
all the logarithmic singularities in the point $X=1$ disappear.
In the same way, spurious singularities in $X=0$ can be eliminated
choosing $C_2=0$.
\smallskip
At this point we are ready to construct the Green function of the
massive scalar fields on $\hyp$. We choose that Green function in such a way
that there is only a logarithmic singularity in $X=0$ on the half-line
$X\ge 0$.
This requirement completely removes the arbitrariness in solving
\prehyp.
Remembering that $X=-y$, the Green function satisfying eq.
\prehyp\ with the desired pole structure is given by:
$$G_\hyp(X)=-{1\over 4\pi}
\fot{-X}\int_1^{-X}dt{[t(1-t)]^{-1}\over [\fot{t}]^2}-$$
\eqn\gonex{-
(1+\lambda)R\fot{-X}\int_1^{-X}dt{[t(1-t)]^{-1}]^{-1}\over [\fot{t}]^2}
\int_0^tds\fot{s}}
{}From the above discussion and from the properties of the
hypergeometric functions, it is clear that
$G_\hyp(X)$
is a well defined Green function, in particular when the integrand
approaches the point $t=1$. Moreover, $G_\hyp(X)$
is regular on the half-line $X>0$, but has a singularity of the kind
$G_\hyp(X)\sim -{1\over 4\pi}
{\rm log}(-X)$ near $X=0$\foot{Remember that the equation of
motion \eqsmot\ contains an overall negative sign.}.
For the sake of completeness,
however, a rigorous proof will be given in the appendix.
Here we only notice that,
using the series representation of the
hypergeometric function \serepr, it is possible to show that
\eqn\integralf{\int_0^tds\fot{s}=tF(1-\beta,\beta;2|t)}
Exploiting the above equation in \gonex,
one can further simplify the expression of
$G_\hyp(X)$ eliminating the double integral in $s$ and $t$.
We remark also that in the limit
$\mu^2=0$ we have $F(1,0;1|X)=1$. Therefore,
when $\lambda=-1$, it is easy to see that $G_\hyp(X)$
becomes, apart from an infinite constant, the
usual scalar Green function of the massless case:
\eqn\massless{\lim_{{\mu^2,\lambda+1
\to 0}\atop {n=2}}{G_\hyp(X)}=-{1\over 4\pi}{\rm log}\left(
-{X\over 1+X}\right)}
Let us notice that the infinite constant is unavoidable in two
dimensions when taking the small mass limit of the massive
propagator. This is due to the unregularized infrared divergencies.
Apart from that, one can easily show that eq. \massless\ is in
agreement with ref. \calwil, where the massless scalar Green function has
been computed on an hyperbolic disk. The latter is related to $\hyp$
only by a conformal transformation (see below).
Moreover, we can also see from eq. \massless\ and from the definition
\anhrat\ of $X$ that the function
${\rm log}\left(
-{X\over 1+X}\right)$ has not only a logarithmic singularity at $z=w$,
but also a logarithmic singularity of the opposite sign at the {\it
image} point $z=\bar w$, as pointed out in ref. \calwil. This singularity,
lying beyond the border $y=0$ of $\hyp$, is harmless and therefore we
have:
$$-{1\over 4\pi}\triangle{\rm log}\left(
-{X\over 1+X}\right)=\delta^{(2)}(z,w)$$
In the limit of large $X$, instead,
the propagator $G_\hyp(X)$ has the following
asymptotic behavior, obtained using the analytic continuation of
the hypergeometric function at infinity:
$$\lim_{X\to\infty}G_\hyp(X)\sim c_1X^{-{1\over 2}-\sqrt{{1\over 4}+\mu^2}}
+c_2R(\lambda+1)$$
In the above equation $c_1$ and $c_2$ are constants depending on $m^2$ and
$k^2$ which can be easily determined from the formulas given in the appendix.
As it is possible to see, in absence of the inhomogeneous term
$(\lambda+1)R$ the fall off of the propagator at infinity increases
exponentially with the mass.
We remember that, from our settings, the point $X=\infty$ corresponds
to the boundary ${\rm Im}z=0$.
Therefore, the vanishing of $G_\hyp(X)$ in $X=\infty$ for $\lambda=-1$
implies the choice of Dirichlet boundary conditions $\left.
\varphi\right|_{\partial M}=0$ on $\hyp$. The same boundary conditions
are also satisfied by the right hand side of eq. \massless.
The fact that the Dirichlet boundary conditions are privileged on
spaces with negative curvature has been pointed out also in ref. \calwil.
\vskip 1cm
\newsec{THE CASE OF THE SPHERE}
\vskip 1cm
Let us now consider the massive scalar fields on the
sphere $\sph$. We choose the metric
$g_{z\bar z}={1\over(1+z\bar z)^2}$, so that the scalar curvature,
defined as $R=g^{z\bar z}\partial_z\partial_{\bar z}
{\rm log}(g_{z\bar z})$, is strictly positive.
The equation of motion analogous to \eqsmot\ becomes in this case:
\eqn\emsphere{\left[-\mti \partial_z\partial_{\bar z}+\mu^2\right]
G_\sph(z,w)=\delta^{(2)}(z,w)-(\lambda+1)R}
In order to explicitly construct the fundamental solution of \emsphere,
it is convenient
to seek for a Green function of the kind $G_\sph(z,w)\equiv G_\sph(X)$,
where $X$ is given now by:
\eqn\xsph{X={(z-w)(\bar z-\bar w)\over(1+z\bar z)(1+w\bar w)}}
Considering a generic function $F(X)$ as an implicit
function of $z,\bar z$, one obtains the following result:
\eqn\srelfund{\mti \partial_z\partial_{\bar z}F(X)=
X(1-X)F^{\prime\prime}(X)+(1-2X)F'(X)+ XF'(X)\delta^{(2)}(z,w)}
As a consequence, the equation of motion \emsphere\ takes the form:
\eqn\shyperg{
-X(1-X)
G^{\prime\prime}_\sph(X)-(1-2X)G'_\sph(X)+
\mu^2G_\sph(X)=\delta^{(2)}(z,w)-
(\lambda+1)R}
This is exactly an hypergeometric equation and therefore
the Green function can be formally derived as in the previous section.
A difference with respect to the case of negative curvature
is however provided by the fact that the parameter $\beta$, depending
on the mass $\mu^2$, is now of the form:
$$\beta={1\over 2}-\sqrt{{1\over 4}-\mu^2}$$
Therefore $\beta$ is allowed to be complex if $\mu^2>1/4$.
This feature has important consequences in the behavior of the propagator
at large values of $X$ as we will see below.
After this remark, we give the explicit form of the propagator on
$\sph$, which can be computed exploiting the same strategy of
the previous case:
$$G_{\sph}(X)=-{1\over 4\pi}
\fot{X}\int_1^{X}dt{[t(1-t)]^{-1}\over [\fot{t}]^2}+$$
\eqn\sgonex{
(1+\lambda)R\fot{X}\int_1^{X}dt{[t(1-t)]^{-1}]^{-1}\over [\fot{t}]^2}
\int_0^tds\fot{s}}
Also when $\beta$ becomes complex, the fact that $1-\beta=\bar \beta$
assures that
$G_{\sph}(X)$ remains real as it should be.
It is also easy to convince oneself that the only singularity of $G_{\sph}(X)$
occurs near the point $X=0$, where $G_{\sph}(X)\sim{\rm log}(X)$.
Moreover, when $\mu^2=0$, $(1+\lambda)R={1\over 4\pi}$ and $n=2$,
$G_{\sph}(X)$ reduces to
the usual Green
function of the massless scalar fields on the sphere, i.e.:
$$\lim_{{m,\lambda\to 0}\atop {n=2}}G_{\sph}(X)={\rm log}(X)$$
Indeed, the right hand side fulfills the well known equation
of the massless Green function on $\sph$:
$$\triangle {\rm log}(X)=\delta^{(2)}(z,w)-{1\over 4\pi}$$
In the limit of large $X$ the behavior of $G_\sph(X)$ is very different
from that of $G_\hyp(X)$.
As a matter of fact, we have at the leading order:
$$G_\sph(X)\sim c'_1X^{-{1\over 2}}{\rm log}(X)+c'_2(\lambda-1)$$
The decreasing at infinity of $G_\sph(X)$ without the
inhomogeneous term proportional to $R(\lambda+1)$ is independent of the mass
term, which contributes only to a complex
phase in the coefficients $c'_1$ and
$c'_2$.
\vskip 1cm
\newsec{THE CASE OF THE HYPERBOLIC DISK}
\vskip 1cm
Finally, we investigate the case in which the topology is given by
a two dimensional disk $\dsk$ of radius $\rho$,
equipped with a metric of the kind
$g_{z\bar z}dzd\bar z={\rho^4dzd\bar z\over (\rho^2-z\bar z)^2}$.
The scalar curvature on $\dsk\otimes \eucl$ is negative:
$R=-{1\over \rho^2}$.
As a matter of fact, the hyperbolic disk $\dsk$ can be obtained from
$\hyp$ after performing the conformal transformation
$z=J(\zeta)$, where $z\in\dsk$, $\zeta\in\hyp$ and $J(\zeta)={i\zeta+
\rho\over \zeta+i\rho}$.
In this way $\dsk$ provides a good test in order to confirm the
results obtained in sections 2 and 3.
Now the equation of motion of the massive scalar fields takes the form:
\eqn\emdisk{\left[-(1-{z\bar z\over\rho^2}
)^2\partial_z\partial_{\bar z}+\mu^2\right]
G_\dsk(z,w)=\delta^{(2)}(z,w)-(\lambda+1)R}
To solve eq. \emdisk\ it is convenient to consider the ansatz $G_\dsk(z,w)
\equiv G_\dsk(X)$, where
\eqn\dskanh{X={\rho^2(z-w)(\bar z-\bar w)\over(\rho^2-z\bar z)(\rho^2-
w\bar w)}}
Repeating the same procedure followed in the previous section, we
arrive at the following equation, valid for a generic differentiable
function F(X):
$$(1-{z\bar z\over \rho^2})^2\partial_z\partial_{\bar z}F(X)=X(1+X)
F^{\prime\prime}(X)+(1+2X)F'(X)+ XF'(X)\delta^{(2)}(z,w)$$
As a consequence, eq. \emdisk\ becomes:
\eqn\dhyperg{
-X(1+X)
G^{\prime\prime}_\dsk(X)-(1+2X)G'_\dsk(X)+\mu^2 G_\dsk(X)=
\delta^{(2)}(z,w)-(\lambda+1)R}
Performing the substitution $y=-X$ as we did for $\hyp$, one gets
the same hypergeometric equation \hyperg. The desired propagator
is therefore given by:
$$G_\dsk(X)=-{1\over 4\pi}
\fot{-X}\int_1^{-X}dt{[t(1-t)]^{-1}\over [\fot{t}]^2}-$$
\eqn\gonex{
(1+\lambda)R\fot{-X}\int_1^{-X}dt{[t(1-t)]^{-1}]^{-1}\over [\fot{t}]^2}
\int_0^tds\fot{s}}
As expected, this
Green function coincides with $G_\hyp(X)$ after performing the
conformal transformation
$J:\dsk\rightarrow\hyp$
introduced above. As a matter of fact, applying
$J$ to $X$ as function of the variables $z$ and $w$ given by \dskanh,
$G_\dsk(X)$ becomes exactly equal to $G_\hyp(X')$, where $X'$
is the  anharmonic ratio \anhrat\ in the variables
$\zeta=J(z)$ and $\omega=J(w)$.
\vskip 1cm
\newsec{THE SCHWINGER MODEL ON 2-D HYPERBOLIC GEOMETRIES}
\vskip 1cm
In this section we consider the massless Schwinger model (or two dimensional
quantum electrodynamics QED$_2$) on
an hyperbolic disk $\dsk$ of radius $\rho$.
The extension to the Poincar\'e upper half plane $\hyp$ can be achieved
performing a conformal
transformation and is straightforward.
The action of the model is given by:
\eqn\schwing{S_{{\rm QED}_2}=\int_\dsk {\rho^4dxdy\over \rho^2-|\vec\xi|^2}
\left[
{1\over 4}F_{\mu\nu}F^{\mu\nu}-\bar \psi
e_\alpha^\mu\gamma^\alpha(\nabla_\mu+ieA_\mu)
\psi\right]}
In the above equation we have exploited the following notations.
The metric is
\eqn\metricnew{g_{\mu\nu}(\vec\xi)=f(\vec\xi)
\delta_{\mu\nu}\qquad\qquad\qquad
\vec\xi=(x,y)}
with
$$f(\vec\xi)={\rho^4\over(\rho^2-|\vec\xi|^2)^2}=
{\rho^4\over(\rho^2-z\bar z)^2}$$
The $e_\alpha^\mu(\vec\xi)$, where
$\alpha,\mu=0,1$, represent the vierbeins and $\nabla_\mu$
denotes the covariant derivative acting on the fermions.
Finally, the $\gamma^\alpha$ are the usual two dimensional $\gamma-$matrices
valid in the flat Euclidean space:
$$\gamma^1=\left(\matrix{0&1\cr 1& 0\cr}\right)$$
$$\gamma^2=\left(\matrix{0&-i\cr i&0\cr}\right)$$
$$\gamma^5\equiv\gamma^3=\left(\matrix{1&0\cr0&-1\cr}\right)$$
In local complex coordinates $z$ and $\bar z$, the Dirac operator $D_\alpha=
e_\alpha^\mu\gamma^\alpha(\nabla_\mu+ieA_\mu)$ has the following components:
\eqn\dopcompone{
D_z=2\left[f^{-{1\over 2}}(\partial_z-ieA_z)+{\bar z\over 2\rho^2}\right]}
\eqn\dopcomptwo{D_{\bar z}=-2\left[f^{-{1\over 2}}
(\partial_{\bar z}-ieA_{\bar z})+
{z\over 2\rho^2}\right]}
To simplify the action \schwing\ it is convenient to decompose the gauge
fields using the Hodge decomposition:
\eqn\hodgedec{A_\mu=i\epsilon_{\mu\nu}\partial^\nu\varphi+\partial_\mu\chi}
where $\varphi$ and $\chi$ are real scalar fields obeying the auxiliary
conditions
$$\int_\dsk d^2\xi\sqrt{g}\varphi(\vec \xi)\ne0\qquad\qquad\qquad
\int_\dsk d^2\xi\sqrt{g}\chi(\vec\xi)\ne 0$$
Now we perform the following transformation on the fermionic fields:
\eqn\chiralone{\psi(\vec\xi)=e^{\iota e\left(\gamma_5\varphi(\vec\xi)+
\chi(\vec\xi)\right)}\psi'(\vec\xi)}
\eqn\chiraltwo{\bar\psi(\vec\xi)=\bar\psi'(\vec\xi)
e^{\iota e\left(\gamma_5\varphi(\vec\xi)-
\chi(\vec\xi)\right)}}
In this way, the massless Schwinger model \schwing\ becomes a free field
theory with an effective action containing an anomalous term.
The latter is a pure quantum effect and can be explained in the path
integral formalism by the noninvariance of the fermionic functional measure
under the chiral transformation \chiralone-\chiraltwo\ \ref\fuji{\prl{
K. Fujikawa}{42}{1979}{1195}; \prd{K. Fujikawa}{21}{1980}{2848};
\prd{K.
Fujikawa}{22}{1980}{1499(E)}.}--\ref\netodas{\prd{J. Barcelos-Neto and
Ashok Das}{33}{1986}{2262}.}. In order to obtain the explicit expression of
the anomalous term, we compute the determinant of the chiral operator
$\ng=e_\alpha^\mu\gamma^\alpha(\nabla_\mu+ieA_\mu)$.
This calculation can be
performed by means of heat kernel techniques.
The case of the Poincar\`e disk
is however exceptional, because the metric $g_{\mu\nu}$ blows up
exactly at the boundary Re$(z)=0$. Most of the scientific literature on
the subject
assumes instead that the metric is finite \chiramath--\msit.
To avoid this difficulty, we will compute the chiral determinant
perturbatively. In {\bf QED}$_2$, in fact,
the one loop radiative correction to the two point function of the
gauge fields is sufficient
in order to determine the exact result \aar.
In other words, we have that:
$${\rm Tr}
\left\{
{\rm ln}
\left[
{
{\rm det}
\left (e^\mu_\alpha\gamma^\alpha(\nabla_\mu+ieA_\mu)\right)\over
{\rm det}\left(e^\mu_\alpha\gamma^\alpha\nabla_\mu\right)}\right]
\right\}=$$
\eqn\anomaly{
{e^2\over 2}\int d^2\xi d^2\xi'\sqrt{g(\vec \xi)}\sqrt{g(\vec\xi')}\langle
\bar\psi(\vec\xi)e^\mu_\alpha(\vec\xi)\gamma^\alpha\psi(\vec\xi)\bar\psi(
\vec\xi')e^\nu_\beta(\vec\xi')\gamma^\beta\psi(\vec\xi')\rangle A_\mu(
\vec\xi)A_\nu(\vec\xi')}
The first ingredient needed in the calculation of the right hand side of eq.
\anomaly\
is the free propagator of the fermionic
fields. In complex notations, see e.g.
\ref\amv{L. Alvarez-Gaum\'e, G. Moore and C. Vafa, {\it Comm. Math.
Phys.} {\bf 106} (1986), 1.}, the two
components of this propagator are given by:
\eqn\propone{\langle \bar\psi_\theta(z,\bar z)\psi_\theta(w,\bar w)\rangle
\equiv S_{\theta\theta}
(z,w)}
\eqn\proptwo{\langle \bar\psi_{\bar\theta}(z,\bar z)\psi_{\bar \theta}
(w,\bar w)\rangle=S_{\bar\theta\bar\theta}
(z,w)}
and are characterized by the fact that,
under a conformal transformation $z\rightarrow z'=z'(z)$, they transform in the
following way:
$$\psi_\theta(z,\bar z)=\psi_{\theta'}(z',\bar z')\left({dz'\over dz}
\right)^{1\over 2}\qquad\qquad\qquad
\bar\psi_\theta(z,\bar z)=\bar\psi_{\theta'}(z',\bar z')\left({dz'\over dz}
\right)^{1\over 2}$$
and
$$\psi_{\bar\theta}(z,\bar z)=\psi_{\bar\theta'}
(z',\bar z')\left({d\bar z'\over d\bar z}
\right)^{1\over 2}\qquad\qquad\qquad
\bar\psi_{\bar\theta}
(z,\bar z)=\bar\psi_{\bar\theta'}(z',\bar z')\left({d\bar z'\over d\bar z}
\right)^{1\over 2}$$
The operators $D_z$ and $D_{\bar z}$ of eqs. \dopcompone\ and
\dopcomptwo\ can be rewritten also in a simplified form, which will be
convenient in future calculations:
$$D_z=2f^{-{3\over 4}}(\partial_z-ieA_z)f^{1\over 4}$$
$$D_{\bar z}=2f^{-{3\over 4}}(-\partial_{\bar z}+ieA_{\bar z})f^{1\over 4}$$
As a consequence we have that $\nabla_z=2f^{-{3\over 4}}(\partial_z)
f^{1\over 4}$, $\nabla_{\bar z}=2f^{-{3\over 4}}(-\partial_{\bar z})
f^{1\over 4}$ and the solutions of the free equations of motion
$$\nabla_zS_{\bar \theta\bar\theta}(z,w)=\delta_{\bar z z}(z,w)\qquad
\qquad\qquad\nabla_{\bar z}S_{\theta\theta}(z,w)=\delta_{\bar z z}(z,w)$$
are simply given by
\eqn\propfin{S_{\theta\theta}(z,w)={1\over \pi}f^{-{1\over 4}}(z,\bar z)
f^{-{1\over 4}}(w,\bar w){1\over z-w}\qquad\qquad
S_{\bar\theta\bar\theta}(z,w)={1\over \pi}f^{-{1\over 4}}(z,\bar z)
f^{-{1\over 4}}(w,\bar w){1\over \bar z-\bar w}}
This is the final expression of the propagator. Image charges are ruled out
by the requirement of covariance under the PSL$(2,{\rm\bf R})$ group of
transformations. For example, the propagator
$$S_{\theta\theta}(z,w)={1\over \pi}f^{-{1\over 4}}(z,\bar z)
f^{-{1\over 4}}(w,\bar w)\left [{1\over z-w}-{1\over \bar z-\bar w}\right]$$
satisfies the above equations of motion but does not transform according
to the rule
$$S_{\theta\theta}(\gamma(z),\gamma(w))=(cz+d)(cw+d)S_{\theta\theta}(z,w)$$
where
$$\gamma=\left[\matrix{a&b\cr c&d\cr}\right]$$
is an element of PSL$(2,{\rm\bf R})$. In real coordinates, $\vec\xi$ and
$\vec\xi'$ we get:
\eqn\propreal{S(\vec\xi,\vec\xi')={1\over 2\pi}
(f(\vec\xi)f(\vec\xi'))^{-{1\over 4}}\gamma_\alpha{(\xi^\alpha-\xi^{\prime
\alpha})\over |\vec\xi-\vec\xi'|^2}}
Using the above propagator, we obtain the exact form of the chiral
determinant:
$${\rm Tr}
\left\{
{\rm ln}
\left[
{
{\rm det}
\left (e^\mu_\alpha\gamma^\alpha(\nabla_\mu+ieA_\mu)\right)\over
{\rm det}\left(e^\mu_\alpha\gamma^\alpha\nabla_\mu\right)}\right]
\right\}=$$
\eqn\anomfinal{-{e^2\over 2\pi}\int d^2\xi d^2\xi'\epsilon^{\mu\nu}
\partial_{\nu_{(\xi)}}A_\mu(\vec\xi){\rm log}|\vec\xi-\vec\xi'|^2
\epsilon^{\rho\sigma}
\partial_{\sigma_{(\xi')}}A_\sigma(\vec\xi')}
Due to the presence of the partial derivatives in $\vec\xi$ and
$\vec\xi'$, the Green function ${\rm log}|\vec\xi-\vec\xi'|^2$
in the right hand side of eq. \anomfinal\ can be replaced by:
\eqn\aaaa{{1\over \triangle}\equiv -{1\over 4\pi}
{\rm log}\left[{|\vec\xi-\vec\xi'|^2\over
\left(1-{\vec\xi^2 \over \rho^2}\right)\left(1-{\vec\xi'^2 \over
\rho^2}\right)}\right]}
where $\triangle\equiv\sqrt{g}\partial_\mu\partial^\mu$ denotes the
Laplacian. Thus eq. \anomfinal\ represents the anomaly of the chiral
Schwinger model in its usual form in curved space-times
\amv.
Decomposing the gauge fields by means of \hodgedec,
we obtain the effective action of the Schwinger model in its final form:
\eqn\actionfinal{S_{{\rm QED}_2}=\int_\dsk d^2\xi{\sqrt{g}\over 2}\left[
\partial_\mu\varphi\left(\triangle-{e^2\over 2\pi}\right)\partial_\mu
\varphi\right]-\int_\dsk d^2\xi\sqrt{g}\left[\bar\psi'\epsilon^\mu_\alpha
\gamma^\alpha\nabla_\mu\psi'\right]}
As we see, the
$\chi$ fields are
completely decoupled and do not contribute as it happens
in the flat case.
The Green function of the scalar fields satisfies the equation:
$$\triangle\left(\triangle-{e^2\over {2\pi}}\right)G(\vec\xi,\vec\xi')=
\delta^{(2)}(\vec\xi,\vec\xi')$$
The solution can be easily obtained from eq. \gonex.
After setting $\lambda=-1$, $X={\rho^2|\vec\xi-\vec\xi'|^2\over
\left(1-{\vec\xi^2 \over \rho^2}\right)\left(1-{\vec\xi'^2 \over
\rho^2}\right)}$ and $\mu^2={e^2\over 2\pi}$,
the result is:
\eqn\scalprop{G(\vec\xi,\vec\xi')={1\over m^2}\left[G_\dsk(X)_{\lambda=-1}
-G_\dsk(X)_{\lambda=-1\atop {m=0}}
\right]}
where $G_\dsk(X)_{\lambda=-1\atop{m=0}}={\rm log}\left(-{X\over 1+
X}\right)$ and
$$G_\dsk(X)_{\lambda=-1}=
-\fot{-X}\int_1^{-X}dt{[t(1-t)]^{-1}\over [\fot{t}]^2}$$
Finally, the Green function of the fermionic fields $\bar\psi$ and
$\psi$ has
been already computed and it is given  by eq. \propreal.
\smallskip
Now we are ready to derive the correlation functions of the Schwinger
model. The most interesting correlators are those involving the original
fermionic fields $\bar\psi$ and $\psi$ before of the transformation
\chiralone-\chiraltwo. As an example, we consider here the
$\bar\psi\psi$ 2--point function:
\eqn\intcalc{\langle\bar\psi(\xiv)\psi(\xiv)\bar\psi(\xiv')\psi(\xiv')\rangle=
\langle\bar\psi'(\xiv)e^{ie\gamma_5\varphi(\xiv)}\psi'(\xiv)
\bar\psi'(\xiv')e^{ie\gamma_5\varphi(\xiv')}\psi'(\xiv')\rangle}
The propagators of the scalar fields $\varphi(\xiv)$ and of the
fermionic fields $\bar\psi$, $\psi$, are given in eqs. \scalprop\ and
\propreal\ respectively. Using these propagators, we obtain from
\intcalc:
\eqn\twopfinal{
\langle\bar\psi(\xiv)\psi(\xiv)\bar\psi(\xiv')
\psi(\xiv')\rangle=
{1\over 2\pi}
(f(\vec\xi)f(\vec\xi'))^{-{1\over 4}}
{
\gamma_\alpha(\xi^\alpha-
\xi^{\prime
\alpha})\over |\vec\xi-\vec\xi'|^2
}
e^{
\left[-4e^2\left(G(\xiv,\xiv)+G(\xiv',\xiv')-2G(\xiv,\xiv')
\right)\right]
}
}
where
$$G(\xiv,\xiv)=G(\xiv',\xiv')=\lim_{X\to0}{1\over
m^2}\left[G_\dsk(X)_{\lambda=1}-G_\dsk(X)_{{\lambda=1}\atop\mu^2=0}\right]$$
It is easy to see that this limit exists and is finite.
Thus, the right hand side of eq. \twopfinal\ has the expected behavior at
short and long distances. When $\xi\sim\xi'$, in fact, it turns out that
$$G(\xi,\xi)+G(\xi',\xi')-2G(\xi,\xi')\sim0$$
and only the free fermionic propagatpr contributes.
At long distances $\xi\rightarrow\infty$, we have instead that the four point
function \twopfinal\ converges to a finite number. To show this,
let us for instance
fix the value of $\xi'$ and study the limit $\xi\rightarrow\infty$ of
\twopfinal.
Clearly
$$\lim_{\xi\to\infty}{f(\xi)^{-{1\over 2}}\over |\xi-\xi'|^2}={\rm
const.}$$
Moreover, it is possible to check from eq. \dskanh\
that in the limit of large $\xi$ the variable $X$ is a finite constant,
depending only on $\xi'$. As a consequence, also the exponent appearing in
eq. \twopfinal\ remains finite at large distances completing our proof.
Finally one has to check the behavior of
$\langle\bar\psi(\xiv)\psi(\xiv)\bar\psi(\xiv')
\psi(\xiv')\rangle$ near the boundary $|z|=\rho$, where $X$ approaches
infinity. Also in this case a straightforward calculation shows
that the 2--point
$\bar\psi\psi$ function converges to a finite result.
\vskip 1cm
\newsec{CONCLUSIONS}
\vskip 1cm
In this paper we have quantized the massive scalar fields and the Schwinger
Model on some relevant examples of two dimensional manifolds with
negative and positive scalar curvature. The related correlation
functions have been explicitly derived in terms of hypergeometric
functions. Despite of the fact that there are many formal analogies
between the case of the complex sphere and the Poincar\'e disk, it
turns out that the behavior of the propagators in the infrared regime
is very different, in particular if $\mu^2\le {1\over 4}$. Moreover,
field theories on hyperbolic manifolds are complicated by the presence
of the boundary. For instance, the calculation of the anomalous chiral
determinant required in order to solve the Schwinger model is mathematically
nontrivial on a Poincar\'e disk. In fact, this is a
noncompact manifold with a metric tensor which diverges when
approaching the boundary. The exact calculation of the chiral
determinant of the Schwinger model could be computed here exploiting
perturbation theory. Our result can shed some light also in similar
problem arising in higher dimensions when hyperbolic manifolds with
constant curvature are considered \cvz--\zerb, \ref\camporesi{R.
Camporesi, {\it Phys. Rep.} {\bf 196} (1990), 1.}.
\smallskip
Finally, it is the first time that the Schwinger model has been
considered also on surfaces with negative curvatures like the
Poincar\'e disk. Since the latter is equivalent to the upper half
plane $\hyp$ up to a conformal transformation, our results are a first
step toward a complete solution of the Schwinger model on a closed and
orientable Riemann surface $\Sigma$. As a matter of fact, a Riemann
surface may be viewed as the ratio $\Sigma={\hyp/\Gamma}$, where
$\Gamma$ is a Fuchsian group (see e. g. \ref\dok{E. D'Hoker and D. H.
Phong, {\it Rev. Mod. Phys.} {\bf 60} (4) (1988), 917.}).
Until now, only the partition function and the generating functional
of the fermionic currents have been nonperturbatively computed on a
Riemann surface \ffscht. In this sense, our calculations can be useful
in at least two  ways. First of all, exploiting the theory of
holomorphic forms on ${\hyp/ \Gamma}$, it is possible to provide a
closed expression for the propagator of the effective mesonic theory
\actionfinal. In \ffscht\ the analogous of this Green function on
$\Sigma$ was in fact given only in terms of an infinite series, supposing
that the mass term $\mu^2={e^2\over 2\pi}$ is small.
Moreover, since the free propagator of the fermionic fields exists in
terms of theta functions \ref\vv{\npb{E. Verlinde and H.
Verlinde}{288}{1987} {357}; \ijmpa{M. Bonini and R.
Iengo}{3}{1988}{841}.}, the idea of computing the chiral determinant
by means of perturbation theory should  work also on Riemann surfaces.
\vskip 1cm
\newsec{ACKNOWLEDGMENTS}
\vskip 1cm
The author wishes to thank G. Cognola, L. Vanzo and S. Zerbini for
many valuable
discussions and fruitful suggestions.
\vskip 1cm
\appendix{A}{}
\vskip 1cm
In this appendix we investigate the behavior of the hypergeometric functions
$\eta_1,\eta_2$ and $\bar \eta$ discussed in section 3.
The hypergeometric formulas are taken from \erdelyi.
The behavior in $y=0$ of $\eta_2(y)$ is provided by the following
expansion:
$$\eta_2(y)=\fot{y}{\rm
log}(y)+\sum\limits_{k=1}^\infty z^k{(1-\beta)_k(\beta)_k\over
(k!)^2}\left[h(k)-h(0)\right]$$
where $h(k)=\psi(1-\beta+k)+\psi(\beta+k)-2\psi(k+1)$ and
$\psi(z)={d\over dz}{\rm log}\Gamma(z)$.
Clearly, in $y=0$ the leading order term is $\eta_2(y)\sim
{\rm log}(y)$.\smallskip
In order to study the hypergeometric functions at the point $y=1$ the
following expansion turns out to be very useful:
$$F(a,b;a+b+l|z)={\Gamma(l)\Gamma(a+b+l)\over
\Gamma(a+l)\Gamma(b+l)}\sum\limits_{n=0}^{l-1}{(a)_n(b)_n\over
(1-l)_nn!}(1-z)^n+$$
\eqn\expyeqone{(1-z)^l(-1)^l{\Gamma(a+b+l)\over\Gamma(a)\Gamma(b)}
\sum k_{n=0}^\infty{(a+l)_n(b+l)_n\over n!(n+l)!}(1-z)^n}
where
$$k_n=\psi(n+1)+\psi(n+1+l)-\psi(a+n+l)-\psi(b+n+l)$$
Moreover $l$ is a nonnegative integer and
the first sum should be set to zero if $l=0$.
Exploiting eq. \expyeqone\ in the case $a=1-\beta$, $b=\beta$ and
$l=0$, it turns out that the leading order behavior of $\eta_1(y)$ at
$y=1$ is
given by:
$$\fot{y}\sim-{\Gamma(1)\over\Gamma(1-\beta)\Gamma(\beta)}{\rm
log}(1-y)+{\rm finite}$$
Consequently, in a neighborhood of $y=1$, setting $y=1-\epsilon$
with $\epsilon$ small, we have:
$$\eta_2(1-\epsilon)\sim {\rm log}(\epsilon)\int^{1-\epsilon}_1
{dt\over 1-t}[{\rm log}(1-t)]^{-2}$$
This shows that the limit in $y=1$ of $\eta_2(y)$ is well
defined and amounts to a constant.
This situation is improved in the case of the function \ypart.
As a matter of fact the integrand in $t$ appearing in
the definition of $\bar\eta(y)$ is the same as that of
$\eta_2(y)$ apart from the multiplication by the function
$tF(1-\beta,\beta;2|t)$.
However, this function vanishes in the neighborhood of the point
$t=1$, cancelling the singularity given by the presence of the factor
$(1-t)^{-1}$.
This can be easily shown using eq. \expyeqone\ with $l=2$, which
yields in $t\sim 1$:
$$tF(1-\beta,\beta;2|t)\sim (1-t){\rm log}(1-t)$$
Finally, it is possible to investigate
the behavior of the propagators near $X=\infty$ by means of the following
formula:
$$\fot{X}={\Gamma(1)\Gamma(2\beta-1)\over\Gamma^2(\beta)}(-1)^{-1-\beta}
z^{\beta-1}F(1-\beta,1-\beta;2-2\beta|{1\over z})+$$$$
{\Gamma(1)\Gamma(1-2\beta)\over\Gamma^2(1-\beta)}(-1)^\beta z^{-\beta}
F(\beta,\beta;2\beta|{1\over z})$$
\listrefs
\end